\documentstyle[fleqn,twoside]{article}

\makeatletter

\@addtoreset{equation}{section}
\makeatother

\def\Tr{\mathop{\rm Tr}\nolimits}

\begin{document}

%%%%%%%%%%%%%%%%%%%%%%%%%%%%%%%%%%%%%%%%%%%%%%%%%%%%%%%%%%%%%%%%%

%%%%%%%%%%%%%%%%%%%%%%%%%%%%%%%%%%%%%%%%%%%%%%%%%%%%%%%%%%%%%%%%%
%%     GC-98.STY:  some definitions and declarations           %%
%%                 for Gravitation & Cosmology                 %%
%%%%%%%%%%%%%%%%%%%%%%%%%%%%%%%%%%%%%%%%%%%%%%%%%%%%%%%%%%%%%%%%%
%%     Copyright:   Kirill Bronnikov, 01.01.1998               %%
%%%%%%%%%%%%%%%%%%%%%%%%%%%%%%%%%%%%%%%%%%%%%%%%%%%%%%%%%%%%%%%%%

\topmargin -8mm
\oddsidemargin -6mm
\evensidemargin -11mm
\textheight 240mm
\textwidth 174mm
\columnsep 8mm
\columnseprule 0.2pt
\emergencystretch=6pt
\mathsurround=1pt
\mathindent=1em
\pagestyle{myheadings}
\newcommand{\bls}[1]{\renewcommand{\baselinestretch}{#1}}
\def\onecol{\onecolumn \mathindent=2em}
\def\twocol{\twocolumn \mathindent=1em}
\def\noi{\noindent}

%% FOR SECTIONING AND HEADINGS:

\renewcommand{\section}{\@startsection{section}{1}{0pt}%
        {-3.5ex plus -1ex minus -.2ex}{2.3ex plus .2ex}%
        {\large\bf\protect\raggedright}}
\renewcommand{\thesection}{\arabic{section}.}
\renewcommand{\subsection}{\@startsection{subsection}{2}{0pt}%
        {-3ex plus -1ex minus -.2ex}{1.4ex plus .2ex}%
        {\normalsize\bf\protect\raggedright}}
\renewcommand{\thesubsection}{\arabic{section}.\arabic{subsection}.}
\renewcommand{\thesubsubsection}%
        {\arabic{section}.\arabic{subsection}.\arabic{subsubsection}.}

\renewcommand{\@oddhead}{\raisebox{0pt}[\headheight][0pt]{%
   \vbox{\hbox to\textwidth{\rightmark \hfil \rm \thepage \strut}\hrule}}}
\renewcommand{\@evenhead}{\raisebox{0pt}[\headheight][0pt]{%
   \vbox{\hbox to\textwidth{\thepage \hfil \leftmark \strut}\hrule}}}
\newcommand{\heads}[2]{\markboth{\protect\small\it #1}{\protect\small\it #2}}
\newcommand{\Acknow}[1]{\subsection*{Acknowledgement} #1}

%% FOR TITLE BLOCK

\newcommand{\Arthead}[3]{ \setcounter{page}{#2}\thispagestyle{empty}\noi
    \unitlength=1pt \begin{picture}(500,40)
        \put(0,58){\shortstack[l]{\small\it Gravitation \& Cosmology,
                        \small\rm Vol. 4 (1998), No. #1, pp. #2--#3\\
        \footnotesize \copyright \ 1998 \ Russian Gravitational Society} }
    \end{picture}          }
\newcommand{\Title}[1]{\noi {\Large #1} \\}
\newcommand{\Author}[2]{\noi{\large\bf #1}\\[2ex]\noindent{\it #2}\\}
\newcommand{\Authors}[4]{\noi
        {\large\bf #1\dag\ #2\ddag}\medskip\begin{description}
        \item[\dag]{\it #3} \item[\ddag]{\it #4}\end{description}}
\newcommand{\Rec}[1]{\noi {\it Received #1} \\}
\newcommand{\Recfin}[1]{\noi {\it Received in final form #1} \\}
\newcommand{\Abstract}[1]{\vskip 2mm \begin{center}
        \parbox{16.4cm}{\small\noi #1} \end{center}\medskip}
\newcommand{\RAbstract}[3]{ {\bf\noi #1}\\ {\bf\noi #2}
        \begin{center}\parbox{16.4cm}{\small\noi #3} \end{center}\bigskip}
\newcommand{\PACS}[1]{\begin{center}{\small PACS: #1}\end{center}}
\newcommand{\foom}[1]{\protect\footnotemark[#1]}
\newcommand{\foox}[2]{\footnotetext[#1]{#2}}
\newcommand{\email}[2]{\footnotetext[#1]{e-mail: #2}}

%%  FOR TEXT, SPACES AND FIGURES

\newcommand{\Ref}[1]{Ref.\,\cite{#1}}
\newcommand{\sect}[1]{Sec.\,#1}
\newcommand{\ssect}[1]{Subsec.\,#1}
\def\ten#1{\mbox{$\cdot 10^{#1}$}}
\def\nq{\hspace*{-1em}}
\def\nqq{\hspace*{-2em}}
\def\nhq{\hspace*{-0.5em}}
\def\nhh{\hspace*{-0.3em}}
\def\cm{\hspace*{1cm}}
\def\inch{\hspace*{1in}}
\newcommand{\Figure}[2]{\begin{figure}
        \framebox[83mm]{\rule{0cm}{#1}}
        \caption{\protect\small #2}\medskip\hrule\end{figure}}
\newcommand{\Theorem}[2]{\medskip\noi {\bf #1. \ }{\sl #2}\medskip}

%% FOR EQUATIONS

\newcommand{\sequ}[1]{\setcounter{equation}{#1}}
\newcommand{\Eq}[1]{Eq.\,(\ref{#1})}
\def\eq{Eq.\,}
\def\eqs{Eqs.\,}
\def\beq{\begin{equation}}
\def\eeq{\end{equation}}
\def\bear{\begin{eqnarray}}
\def\al{&\nhq}
\def\lal{&&\nqq {}}               % left alignment
\def\bearr{\begin{eqnarray} \lal}
\def\ear{\end{eqnarray}}
\def\earn{\nonumber \end{eqnarray}}
\def\dst{\displaystyle}
\def\tst{\textstyle}
\newcommand{\fracd}[2]{{\dst\frac{#1}{#2}}}
\newcommand{\fract}[2]{{\tst\frac{#1}{#2}}}
\def\nn{\nonumber\\ {}}
\def\nnv{\nonumber\\[5pt] }
\def\nnn{\nonumber\\ \lal }
\def\nnnv{\nonumber\\[5pt] \lal }
\def\yy{\\[5pt] {}}
\def\yyy{\\[5pt] \lal }
\def\eql{\al =\al}

\def\eqdef{\stackrel{\rm def}=}
\def\e{{\,\rm e}}
\def\d{\partial}
\def\re{\mathop{\rm Re}\nolimits}
\def\im{\mathop{\rm Im}\nolimits}
\def\arg{\mathop{\rm arg}\nolimits}
\def\tr{\mathop{\rm tr}\nolimits}
\def\sign{\mathop{\rm sign}\nolimits}
\def\diag{\mathop{\rm diag}\nolimits}
\def\dim{\mathop{\rm dim}\nolimits}
\def\const{{\rm const}}
\def\Half{{\dst\frac{1}{2}}}
\def\half{{\tst\frac{1}{2}}}
\def\then{\ \Rightarrow\ }
\def\chg{\ \leftrightarrow\ }
\newcommand{\aver}[1]{\langle \, #1 \, \rangle \mathstrut}
\def\DAL{\raisebox{-1.6pt}{\large $\Box$}\,}
\newcommand{\vars}[1]{\left\{\begin{array}{ll}#1\end{array}\right.}
\newcommand{\lims}[1]{\mathop{#1}\limits}
\newcommand{\limr}[2]{\raisebox{#1}{${\lims{#2}}$}}
\def\suml{\sum\limits}
\def\intl{\int\limits}
\def\wider{\vphantom{\int}}
\def\wideup{\vphantom{\intl^a}}

%%%%%%%%%%%%%%%%%%%%  END OF GC-98.STY  %%%%%%%%%%%%%%%%%%%%

\heads
{L.N. Granda and S.D. Odintsov}
{Effective Average Action and Nonperturbative
Renormalization Group Equation }
%in Higher Derivative Quantum Gravity}

\twocolumn[
\Arthead{2 (14)}{1}{10}
\bigskip

\Title
{EFFECTIVE AVERAGE ACTION \yy
AND NONPERTURBATIVE RENORMALIZATION GROUP EQUATION \yy
IN HIGHER DERIVATIVE QUANTUM GRAVITY}

\Authors {L.N. Granda}{and S.D. Odintsov}
{Departamento de F\'\i sica, Universidad del Valle,
     A.A. 25360, Cali, Colombia}
{Departamento de F\'\i sica, Universidad del Valle,
     A.A. 25360, Cali, Colombia\\
and Tomsk Pedagogical University, 634041 Tomsk, Russia\\
and Department of Physics, Hiroshima University, Higashi-Hiroshima, Japan}

\Rec{25 November 1997}

\Abstract
{We study the exact renormalization group (RG) in $R^2$-gravity
in the effective average action formalism using the background field method.
The truncated evolution equation (where truncation is made to
low-derivatives functionals space) for such a theory in a de Sitter
background leads to a set of nonperturbative RG equations for
cosmological and gravitational coupling constants. The gauge dependence
problem is solved by working in the physical Landau-DeWitt gauge
corresponding to gauge-fixing independent effective action. Approximate
solution of nonperturbative RG equations reveals the appearence of
antiscreening or screening behaviour of Newtonian coupling, depending on
the higher-derivatives coupling constants. The existence of unstable UV
fixed points is also mentioned.}

\vspace{53mm}

% \RAbstract
% {dd%*b("-.% a`%$-%% $%)ab"(% ( -%/%`bc`! b("-.% `%-.`,#`c//.".%
%  c` "-%-(% " *" -b.".) #` "(b f(( a "kah(,( /`.('".$-k,(}
% {.^M. ` -$ , .. $(-f."}
% {aa+%$c%bao b.g- o `%-.,#`c//  () " $R^2$-#` "(b f(( " d.`, +(',%
% mdd%*b("-.#. $%)ab"(o a /.,.iln ,%b.$  d.-.".#. /.+o. a%g%--.% c` "-%-(%
% m".+nf(( (/`( ca%g%-(( * /`.ab` -ab"c dc-*f(.- +." a -('h(,( /`.('".$-k,()
% $+o $ --.) b%.`(( /`(".$(b * a(ab%,% -%/%`bc`! b("-ke -c` "-%-() $+o
% *.a,.+.#(g%a*.) ( #` "(b f(.--.) /.ab.o--ke. `.!+%,  * +(!`."*( `%h %bao
% '  ag%b (a/.+l'." -(o d('(g%a*.) * +(!`."*(  -$ c-%(bb ,
% a..b"%bab"cni%) * +(!`.".g-.--%' "(a(,.,c mdd%*b("-.,c $%)ab"(n.
% `(!+(&%--.% `%h%-(% -%/%`bc`! b("-.#. -c` "-%-(o .!- `c&(" %b /.o"+%-(%
%  -b(m*` -(`." -(o (+( m*` -(`." -(o -lnb.-."a*.#. "' (,.$%)ab"(o "
% ' "(a(,.ab( .b *.-ab -b a"o'( /`( "kah(e /`.('".$-ke. /.,(- %bao
% aci%ab"." -(% -%cab.)g("ke  d(*a(`." --ke b.g%*.  }

] %%%%%%%%%%%%%%%%%%%%%%
\email 1 {granda@quantum.univalle.edu.co}
\email 2 {odintsov@quantum.univalle.edu.co\\ {\it or\/} sergei@ecm.ub.es}

\section{Introduction}

It is well-known that Einstein quantum gravity (QG) is not renormalizable
\cite{dn}. There are QG models which represent extensions of Einstein
gravity. One of them, the so-called $R^2$-gravity (see \cite{bos} for an
introduction and review) is multiplicatively renormalizable. However, it is
most probably a non-unitary theory, at least in the perturbative approach.
In the situation when consistent quantum gravity (QG) is unknown it is
quite reasonable to study the existing gravitational theories as effective
theories.  This gives one the possibility of estimating QG manifestations at
low energy scales.

In such a way one can reduce QG to a simpler theory
described by some type of scalar Lagrangian \cite{a,t}. Those models are
useful for describing QG in the far infrared domain (at large distances)
\cite{a,t}.

One can consider another approach. Let us take non-renormalizable Einstein
gravity and work with it as with a usual non-renormalizable effective
field theory. Then the calculation of quantum corrections is still
possible. In such a way, quantum corrections to the Newtonian coupling
constant and to the Newtonian potential have been estimated \cite{dhm}.

Finally, one can apply the exact RG \cite{ww} in the study of
non-renormalizable theories. There has been recently much activity in
studying different theories (mainly scalar ones) using the
non-perturbative RG (for a list of recent papers see \cite{bam} and
references therein).  Using an average effective action and the background
field method, a non-perturbative RG study of Einstein quantum gravity was
recently presented \cite{mr}. The RG equation (or evolution equation)
governing the evolution of the effective action from a scale
$\Lambda_{\rm cut-off}$ where theory is well-defined to smaller scales
$k<\Lambda_{\rm cut-off}$ has been constructed in  Ref.\,\cite{mr}.  Its
gauge dependence has been investigated in \cite{fo}. It has been shown
\cite{fo} that in the physical gauge (the Landau-DeWitt gauge) the
Newtonian coupling constant shows antiscreening.

In the present paper we formulate the non-pertur\-bative RG equation
(evolution equation) in higher derivative QG (for a review and list of
references see \cite{bos}). It is widely known that such a theory, being
multiplicatively renormalizable and asymptotically free, has a perturbatively
non-unitary S-matrix. Nevertheless, such a theory has a lot of
applications. For example, it may lead to more or less succesful inflation
\cite{sta}. Attempts to construct supersymmetric
generalizations of $R^2$-gravity have been recently made \cite{ov,hn}.

We consider higher-derivative QG as an effective theory, so issues of
renormalizability or (non)unitarity are not important for us. We adopt the
formalism  of Refs.\,\cite{mr,fo} in such a theory and construct the
scale-dependent gravitational average action $\Gamma_k[g_{\mu\nu}]$ in the
background field formalism. A truncated evolution equation is obtained.
We work in the physical gauge corresponding to a gauge-fixing independent
effective action. Note that we make truncation of the average effective
action to the space of low derivatives functionals only. Even in such
a simplified variant (where higher-derivative couplings may be considered as
free parameters) the calculation of nonperturbative RG equations is very
complicated.

The paper is written as follows. In the next section we
a give very brief review of the evolution equation (for more details see
\cite{mr,fo}) and our truncation. \sect 3 is devoted to the calculation of
the one-loop effective action in $R^2$-gravity in the De Sitter background.
Such an evaluation is presented in two cases: (a) a convenient effective
action in a one-parameter-dependent gauge and (b) a gauge-fixing-independent
effective action.  The results of the above calculation are used to
obtain the effective average action in the background field formalism for
these two cases. (In other words, we obtain the r.h.s. of the evolution
equation in the De Sitter background). In \sect 4 we perform an explicit
truncation of the evolution equation and obtain nonperturbative RG
equations for the gravitational and cosmological coupling constants. In
order to avoid the gauge dependence problem we work there in the
gauge-fixing independent EA formalism (see \cite{v} for an introduction).
The critical points of the RG equations for the Newtonian and cosmological
couplings and running Newtonian coupling are discussed in \sect 5.
Finally, some remarks are given in the conclusion.

\section{Evolution equation for average effective action}

We will start from a short introduction to the average effective action
approach in quantum gravity. We follow mainly \Ref{mr} where more
details are presented.

The basic elements of the approach are:

\medskip\noi{\bf 1.}
The background field method \cite{bos} implying that
\beq
	g_{\mu\nu}=\bar{g}_{\mu\nu}+h_{\mu\nu}         \label{(2.1)}
\eeq
where $\bar{g}_{\mu\nu}$ is the background metric and $h_{\mu\nu}$ is a
quantum fluctuation.

\medskip\noi{\bf 2.}
A scale-dependent generating functional for the connected Green functions
\bearr
	W_k [t^{\mu\nu},\sigma^{\mu},\bar{\sigma}_{\mu};
		\beta^{\mu\nu},\tau_{\mu};\bar{g}_{\mu\nu}]
\nnn
	= \int Dh_{\mu\nu}DC^{\mu}D\bar{C}^{\nu}
		\exp\{-S[\bar{g}+h]-S_{\rm gf}[h;\bar{g}]
\nnn
	-S_{gh}[h,C,\bar{C};\bar{g}]
          -\Delta_kS[h,C,\bar{C};\bar{g}]-S_{\rm source}\}.   \label{(2.2)}
\ear

\noi
Let us give a description of the quantities which enter into \eq (2.2).
$S[\bar{g}+h]$ is the classical action of gravity theory under discussion;
$S_{\rm gf}$ denotes the gauge-fixing term. As we will be interested in
$R^2$-gravity, we suppose that $S_{\rm gf}$ may be of the fourth order in
the derivatives. The set of ghosts $C,\bar{C}$ includes all ghosts in the
theory (in $R^2$-gravity we have an extra ghost, the so-called third ghost).
Finally, $\Delta_k S$ is the infrared (IR) cut-off for the gravitational
field and ghosts.  An introduction to the present formalism of studying the
average effective action has been presented in all detail in \cite{mr}, so
we will not present more details of it here. $S _{\rm source}$ in (2.2) is the
standard action describing the coupling of the gravitational field and
ghosts with the sources $t^{\mu\nu}$, $\sigma^{\mu}$, $\bar{\sigma}_{\mu}$.

Performing a Legendre transform of $W_k$
to get the average effective action $\Gamma_k[g,\bar{g}]$, we can obtain the
truncated evolution equation for $\Gamma_k$ (see \cite{mr} for more details)
\bearr
\nq
\d_t\Gamma_k[g,\bar{g}]=\Half \Tr\Bigl[\left(\! A\Gamma_k^{(2)}
	[g,\bar{g}]+R_k^{\rm grav}[\bar{g}]\right)^{\!-1}
			\nhh \d_tR_k^{\rm grav}[\bar{g}]\Bigr]
\nnn
	- \sum_i c_i\Tr\left[\left(-M_i[g,\bar{g}]+R_{ki}^{\rm gh}
       	[\bar{g}]\right)\d_tR_{ki}^{\rm gh}[\bar{g}]\right]   \label{(2.3)}
\ear
where $t=\ln k$, $k$ is the nonzero momentum scale, $A$ is some constant
which depends on the model under discussion (for Einstein gravity
$A=\kappa^2$), $R_k$ are cut-offs, $c_i$ are the weights for ghosts. For the
case of $R^2$-gravity we have the Fadeev-Popov ghost with $c_{\rm FP}=1$ and
the third ghost with weight $c_{\rm TG}=1/2$, and, of course, $M_{\rm FP}$
and
$M_{\rm TG}$ are usually known. $\bar{g}_{\mu\nu}$ is the background metric
and $g_{\mu\nu}=\bar{g}_{\mu\nu}+h_{\mu\nu}$ where $h_{\mu\nu}$ is the
quantum field.  $\Gamma_k^{(2)}$ is the Hessian of $\Gamma_k[g,\bar{g}]$
with respect to $g_{\mu\nu}$ at fixed $\bar{g}_{\mu\nu}$.

The next step is to specify the truncated evolution equation for the theory
under study. We start from the $R^2$-gravity (3.1) at the UV
scale $\Lambda_{\rm cut-off}$ and evolve it down to smaller scales
$k<<\Lambda_{\rm cut-off}$.  We use the truncation where the coupling
constants are replaced by the $k$-dependent functions (see (3.1))
\bearr
	\kappa^2\to  Z_{Nk}^{-1}\kappa^2  ,                  \cm
	\frac{1}{f^2}\to Z_{Nk}\frac{1}{f^2}  ,              \nnn
	\frac{1}{\nu^2}\to Z_{Nk}\frac{1}{\nu^2}  ,          \cm
	\Lambda\to  \bar{\lambda}_k.            \label{(2.4)}
\ear
 Note that we do not write explicitly the $k$-dependence for
 the higher-derivatives coupling constants because we will be restricted
 here only to lower-derivative terms (i.e. up to the linear curvature term).
 Then, in such an approach (subreduction of the full set of RG equations),
 the higher-derivative coupling constants may be considered as free
 parameters of the theory.

 Then, closely following the arguments of Ref.\,\cite{mr}, we get
 (keeping only low-derivative terms)
\beq
 \Gamma_k[g,g]=2\kappa^2Z_{Nk}\int d^4x
 \sqrt{g}\left[-R(g)+2\bar{\lambda}_k\right]. \label{(2.5)}
\eeq
 The ghost term disappears after choosing $\bar{g}_{\mu\nu}=g_{\mu\nu}$.
 Projecting the evolution equation on the space with low-derivatives terms,
 one gets the left-hand side of the truncated evolution equation (2.3) as
 follows
\bearr
 \d_t\Gamma_k[g,g]        \nnn
 	=2\kappa^2\int d^4x
     \sqrt{g}[-R(g)\d_tZ_{Nk}+2\d_t(Z_{Nk}\bar{\lambda}_k)]. \label{(2.6)}
\ear
 The initial conditions for $Z_{Nk}, \lambda_k$ are choosen in the same way
 as in \cite{mr}. The right-hand side of the truncated evolution equation
 (2.3) will be defined in the next section, following similar one-loop
 arguments. We have to note only that, unlike the Einstein gravity,
the projectors should include the coupling constants. We do not give more
details as they are very similar to those discussed in \cite{mr}.

\section{One-loop effective action and
effective average action in $R^2$-gravity}

In this section we study  the one-loop effective action and the average
effective action in higher-derivative quantum gravity (for a review see
\cite{bos} and references therein).

The classical action in Euclidean notations has the following form:
\bearr
 S=\int d^4x \sqrt{g} \biggl\{\epsilon
 	R^*R^*+\frac{1}{2f^2}C_{\mu\nu\alpha\beta}C^{\mu\nu\alpha\beta}
\nnn
 	-\frac{1}{6\nu^2}R^2-2\kappa^2 R+4\kappa^2 \Lambda\biggr\}
	\label{(3.1)}
\ear
 where  $R^*R^*=\frac{1}{4}\epsilon^{\mu\nu\alpha\beta}
\epsilon_{\lambda\rho\gamma\delta}R^{\lambda\rho}_{\mu\nu}
 R^{\gamma\delta}_{\alpha\beta}$, $C_{\mu\nu\alpha\beta}$ is the Weyl tensor,
 $\kappa^{-2}=32\pi \bar{G}$ is the Newtonian coupling constant, $\epsilon,
 f^2, \nu^2$ are the gravitational coupling constants related to the
 higher-derivative  terms in (3.1). It is quite well-known that the theory
 with the action (3.1) is multiplicatively renormalizable and asymptotically
 free (see \cite{bos} for a review).

 Our first purpose will be to calculate the one-loop effective action for
 the theory with action (3.1) on de Sitter background:
\bear
	R_{\mu\nu\alpha\beta}\eql \frac{1}{12}
	\left(g_{\mu\alpha}g_{\nu\beta}-g_{\mu\beta}g_{\nu\alpha}\right)R,
\nn
	R_{\mu\nu}\eql \frac{1}{4}g_{\mu\nu}R.  \label{(3.2)}
\ear

 We work in the usual background field method \cite{bos}, where the
 background field is given by the de Sitter metric,
\beq
g_{\mu\nu}\to  \bar{g}_{\mu\nu}+h_{\mu\nu}            \label{(3.3)}
\eeq
 and $h_{\mu\nu}$ is the quantum gravitational field.

 We will be interested in the calculation of the effective action in
 a parameter-dependent gauge. Then, even in the case of Einstein gravity
 \cite{ft,gp} it is known that one has to make a change of the quantum
 fields:
\bearr
	h_{\mu\nu}=\bar{h}_{\mu\nu}^{\bot}+2\nabla_{(\mu}
	\xi_{\nu)}+\frac{1}{4}g_{\mu\nu}h-\frac{1}{4}g_{\mu\nu}\DAL\sigma ,
\nnn
	h=h^{\mu\nu}g_{\mu\nu} , \cm \nabla^{\mu}\xi_{\mu}^{\bot}=0,
     \cm   \nabla^{\mu}h_{\mu\nu}^{\bot}=0 ,
\nnn
 \xi_{\mu}=\xi_{\mu}^{\bot}+\Half \nabla_{\mu}\sigma , \cm
	 \bar{h}_{\mu\nu}^{\bot}g^{\mu\nu}=0. 		\label{(3.4)}
\ear
 Clearly the transformation (3.4) induces a nontrivial Jacobian which
 should be taken into account in calculating  the one-loop
 effective action.

The second variation of the classical action (3.1) in terms
of the variables (3.4) is written as follows \cite{av}:
\bearr
     \delta^2 S=\int d^4x  \sqrt{g} \biggl\{\frac{1}{4f^2}\bar{h}^{\bot}
\nnn\nhq
 	\times \biggl[\Delta_2
     \Bigl(m_2^2+\frac{f^2{+}\nu^2}{3\nu^2}R\Bigr)
	\Delta_2\Bigl(\frac{R}{6}\Bigr)
     		+\Half m_2^2(R{-}4\Lambda)\biggr]\bar{h}^{\bot}
\nnn\qquad
	-\frac{3}{32\nu^2}\biggl[(h-\DAL \sigma)
       \biggl(\Delta_0(m_0^2)\Delta_0(-R/3)
\nnn \cm \cm
      +\frac{1}{3}m_0^2(R-4\Lambda)\biggr)(h-\DAL\sigma)
\nnn \cm \cm
	-\frac{2}{3}m_0^2(R-4\Lambda)(h-\DAL\sigma)\Delta_0(0)\sigma
\nnn \cm \cm
	-\frac{2}{3}m_0^2(R-4\Lambda)\sigma\Delta_0(0)\Delta_0
      	                 (-R/2)\sigma\biggr]
\nnn \cm
	+2\kappa^2(R-4\Lambda)\epsilon^{\bot}\Delta_1
                       (-R/4)\epsilon^{\bot}\biggr\}   \label{(3.5)}
\ear
 where $m_2^2=2\kappa^2 f^2$, $m_0^2=2\kappa^2 \nu^2$.
 In accordance with \cite{ft} the constrained differential operators are
 introduced:
\bear
	\Delta_0(X)\phi \eql (-\DAL+X)\phi,
\nnv
    \Delta_{1\mu\nu}(X)\xi^{\nu\bot} \eql (-\DAL_{\mu\nu}
                       +g_{\mu\nu}X)\xi^{\nu\bot} ,
\nnv
	\Delta_{2\alpha\beta}^{\mu\nu}(X)\bar{h}_{\mu\nu}^{\bot}
    \eql (-\DAL_{\alpha\beta}^{\mu\nu}+\delta_{\alpha}^{\mu}
    	     \delta_{\beta}^{\nu}X)\bar{h}_{\mu\nu}^{\bot}.  \label{(3.6)}
\ear
\onecol
\noi
At the next step one can choose the gauge fixing term as follows:
\beq
     S_{\rm GF}=\Half \int d^4x \sqrt{g}\chi_{\mu}H^{\mu\nu}\chi_{\nu}
		\label{(3.7)}
\eeq
where
\bear
	\chi_{\mu} = -2\nabla_{\nu}\left\{h^{\nu}_{\mu}
		-\frac{1}{4}(1-K)\delta_{\mu}^{\nu}h\right\} ,
\inch
	H^{\mu\nu} = \frac{1}{4\alpha^2}\left\{g^{\mu\nu}
		 \left(-\DAL+\frac{R}{4}\right)\right\}  \label{(3.8)}
\ear
and $\alpha^2$, $K$ are gauge parameters.

The general expression for the one-loop EA is given by
\bear
	\Gamma=S+\Half \ln\det(\delta^2S+S_{\rm GF})
	-\Half \ln\det H^{\mu\nu}-\ln\det M^{\mu\nu} \label{(3.9)}
\ear
 where the standard ghost operator $M^{\mu\nu}$ is calculated with the help
 of $\chi_{\mu}$ as follows:
\beq
 	M_{\mu\nu}=2\left\{g_{\mu\nu}\Bigl(-\DAL-\frac{R}{4}\Bigr)
 	+\Half (K{-}1)\nabla_{\mu}\nabla_{\nu}\right\} \label{(3.10)}
\eeq
 and the third ghost operator $H^{\mu\nu}$ is given in (3.8). Note that when
 the operators $H^{\mu\nu}$, $M^{\mu\nu}$ are written, the properties of
 the de Sitter background (3.2) are taken into account.

 We follow in the evaluation of the one-loop  EA the results of \Ref{av}
 where this calculation was performed in a more complicated six-parametric
 gauge. For our purposes, the only case $\alpha^2=0$, $K$ being arbitrary
 (a Landau-DeWitt type gauge), will be of interest.

Taking into account the ghost operators and the Jacobian of the variables
change (3.4), we obtain \cite{av}
\bearr
 	\Gamma^{(1)}=\Half \ln\det\biggl[
 	 \Delta_2\left(\frac{R}{6}\right)\Delta_2\left(m_2^2
 		+\frac{f^2+\nu^2}{3\nu^2}R\right)
		+\Half m_2^2(R-4\Lambda)\biggr]
	-\Half \ln\det\Delta_1\left(-\frac{R}{4}\right)
\nnn \cm
	+\Half \ln\det \biggl[\Delta_0^2
	\left(\frac{R}{K-3}\right)\Delta_0(m^2)
	+m_0^2 \frac{K^2-3}{(K-3)^2}\Delta_0\left(\frac{R}{K^2-3}\right)
		(4\Lambda-R)\biggr]
	-\ln\det\Delta_0\left(\frac{R}{K-3}\right)	   \label{(3.11)}
\ear
 	Let us present \eq (3.11) in the form of gravitational and ghost
 	contributions:
\bearr
   \Gamma^{(1)}_{\rm grav}=\Half \ln\det\biggl[\Delta_2\left(\frac{R}
	{6}\right)\Delta_2\left(m_2^2+\frac{f^2+\nu^2}{3\nu^2}R\right)
  	+\Half m_2^2(R-4\Lambda)\biggr]+\Half \ln\det\Delta_1
		\left(-\frac{R}{4}\right)
  +\Half \ln\det\Delta_1\!\left(\frac{R}{4}\right)
\nnn\cm
  +\Half \ln\det
	\biggl[\Delta_0^2\left(\frac{R}{K{-}3}\right)\Delta_0(m_0)^2
  +m_0^2\frac{K^2-3}{(K-3)^2}\Delta_0\left(\frac{R}{K^2-3}\right)
	(4\Lambda-R)\biggr]
	     +\Half \ln\det\Delta_0(0),  \label{(3.12)}
\\ \lal
    \Gamma^{(1)}_{\rm ghost}
    	=-\Half \ln\det H^{\mu\nu}-\ln\det M^{\mu\nu}
\nnn \inch
  = -\Half \ln\left[\det\Delta_1\left(\frac{R}{4}\right)\det\Delta_0(0)
	\right]
     -\ln\left[\det\Delta_1\left(-\frac{R}{4}\right)\det\Delta_0
	\left(\frac{R}{K-3}\right)\right]                      \label{(3.13)}
\ear
 where $\Gamma^{(1)}=\Gamma^{(1)}_{\rm grav}+\Gamma^{(1)}_{\rm ghost}$. The
 gauge dependence of the one-loop effective action is clearly seen in
 \eqs (3.11)--(3.13).

 In order to avoid the explicit gauge dependence one can work with the
 gauge-fixing independent EA (for an introduction see \cite{bos,v}). An
 explicit calculation has been done in \Ref{av} with the following result:
\bearr
	\Gamma^{(1)\rm GFI}_{\rm grav}
	=\Half \ln\det \biggl[\Delta_2\left(\frac{R}{6}
	\right)\Delta_2\left(m_2^2+\frac{f^2+\nu^2}{3\nu^2}R\right)
  	+\frac{1}{2K}m_2^2(R-4\Lambda)\biggr]
	+\Half \ln\det
		\Delta_1\left(-\frac{R}{4}\right)
\nnn \inch
	+\Half \ln\det\Delta_1 \left(\frac{R}{4}\right)
	+\Half \ln\det \biggl[\Delta_0\left(\frac{R}{K-3}\right)
	\Delta_0\left(m_0^2\right)
	-\frac{1}{K-3}m_0^2\left(R-4\Lambda\right)\biggr]
\nnn \inch\inch\inch
	+\Half \ln\det\Delta_0(0)+\Half \ln\det\Delta_0
	\left(\frac{R}{K-3}\right)                             \label{(3.14)}
\ear
 where the parameter $K$ is fixed: $K=3f^2/(f^2+2\nu^2)$. The
 ghost contribution in the gauge-fixing independent EA formalism is given
 again by (3.13) with $K$ fixed as above. Hence, we also found the one-loop
 gauge-fixing independent EA. The use of such EA solves the problem of
 gauge dependence of the EA (for a discussion of the dependence of the
 gauge-fixing independent EA on the metric in the space of fields in quantum
 gravity see \cite{so}). Note that the gauge-fixing independent one-loop EA
 in Einstein quantum gravity in a constant-curvature space
 (a background like the de Sitter space) has been discussed in \Ref{bko},
 (see \cite{bos} for a review).

 Our final goal is related to a study of the truncated evolution
 equation. A necessary step in such a study is the expansion of the average
 effective action in powers of the curvature. Effectively, one
 should use the one-loop effective action in such an expansion.

 However, the transition to constrained differential operators in accordance
 with (3.4) introduces additional zero modes. This leads to a wrong
 answer when we expand the determinants of the constrained operators in
 powers of the curvature. Therefore it is better to represent the effective
 action in terms of unconstrained operators. It could be done with the help
 of the following relations \cite{ft}:
\bearr
	\det\Delta_V(X)\equiv\det(-\DAL+X)_V
	=\det\Delta_1(X) \det\Delta_0\left(X-\frac{R}{4}\right),
\nnn
	\det\Delta_T(X)\equiv\det\left(-\DAL+X\right)_T
	=\det\Delta_2(X)
	\det\Delta_1(X-\fract{5}{12}R)\det\Delta_0(X-\fract{2}{3}R).
							    \label{(3.15)}
\ear
 The operators from the left-hand side are unconstrained. Note also that in
 order to apply the relations (3.15) we should also rewrite the
 higher-derivative operators in terms of low-derivative (second order)
 ones.

 For simplicity, we consider below only the one-loop EA (3.14),
(3.13). This EA actually describes two cases: for $K=1$ it coincides with
the standard EA (3.12), (3.13) in the gauge $K=1$ and for
$K={3f^2}/(f^2+2\nu^2)$ it describes the gauge-fixing-independent EA.

First of all, we rewrite the ghost contribution (3.13) in terms of
unconstrained operators:
\beq
\Gamma^{(1)}_{\rm ghost}=-\Half \ln\det\Delta_V\left(\frac{R}{4}\right)
-\ln\left[\det\Delta_V\left(-\frac{R}{4}\right)\det\Delta_0
\left(\frac{R}{K-3}\right)\det^{-1}\Delta_0\left(-\frac{R}{2}\right)\right]
							\label{(3.16)}
\eeq
Hence the ghost part is expressed in terms of unconstrained operators.

For the gravitational part we get
\bearr
\Gamma^{(1)GFI}_{\rm grav}=\Half \ln\det\left\{\frac{\Delta_T
\left(\frac{b+\sqrt{b^2-4c}}{2}\right)}{\Delta_V
\left(\frac{b+\sqrt{b^2-4c}}{2}-\frac{5}{12}R\right)}\right\}
+\Half \ln\det\left\{\frac{\Delta_T
\left(\frac{b-\sqrt{b^2-4c}}{2}\right)}{\Delta_V
\left(\frac{b-\sqrt{b^2-4c}}{2}-\frac{5}{12}R\right)}\right\}
\nnn \inch
	+\Half \ln\det\left\{\frac{\Delta_V(-R/4)}
	{\Delta_0(-R/2)}\right\}+\Half
	\ln\det\Delta_V\left(\frac{R}{4}\right)
\nnn  \inch
	+\Half \ln\det\left[\Delta_0\left(\frac{R}{K-3}\right)
	\Delta_0(m_0^2)-\frac{1}{K-3}m_0^2\left(R-4\Lambda\right)\right]
	+\Half \ln\det\Delta_0\left(\frac{R}{K-3}\right)
	 						\label{(3.17)}
\ear
	where
\[
	b=\frac{R}{6}+m_2^2+\frac{f^2+\nu^2}{3\nu^2}R,
\inch
	c=\frac{R}{6}\left(m_2^2+\frac{f^2+\nu^2}{3\nu^2}R\right)
	+\frac{1}{2K}m_2^2\left(R-4\Lambda\right)
\]
 Recall that for $K=1$, (3.16) plus (3.17) gives the standard
 one-loop EA in the gauge $K=1$. However, now this EA is expressed in terms
 of unconstrained differential operators.

 Now we can write the average
 effective action in the theory. First of all, to write the evolution
 equation we have to include the cut-off term $\Delta_kS$. In other
 words, in the calculation of $W_k=\ln Z_k$ in the exponent of the path
 integrand we have to consider not only $\Gamma^{(2)}_{k\ \rm grav}$ ,
 $\Gamma^{(2)}_{k\ \rm gh}$ and the ghost term, but also $\Delta_kS$.

 The coefficients $Z_k^{\rm grav}$ and $Z_k^{\rm gh}$ should be chosen so
 that the kinetic and cut-off terms combine to $-\DAL+
 k^2R^{(0)}(-\DAL/k^2)$ for every degree of freedom. Here
$R^{(0)}$ is a dimensionless cut-off function. As in the case of pure
Einstein gravity \cite{mr,fo}, all renormalization effects of ghosts are
neglected.

Hence the effective average action may be written in the form
(see \eqs.(3.16) and (3.17))
\bearr
 \nhq\Gamma_k[g,g]=\Half
 \Tr_T\ln\left\{Z_{Nk}\left(-\DAL+\frac{b+\sqrt{b^2-4c}}{2}+k^2R^{(0)}\right)
 \right\}
\nnn \cm
 +\Half
 \Tr_T\ln\left\{Z_{Nk}\left(-\DAL+\frac{b-\sqrt{b^2-4c}}{2}+k^2R^{(0)}\right)
 \right\}
\nnn \cm
 -\Half
 \Tr_V\ln\left\{Z_{Nk}\left(-\DAL+\frac{b+\sqrt{b^2-4c}}{2}
 -\frac{5}{12}R+k^2R^{(0)}\right)\right\}
\nnn \cm
 -\Half
 \Tr_V\ln\left\{Z_{Nk}\left(-\DAL+\frac{b-\sqrt{b^2-4c}}{2}
 -\frac{5}{12}R+k^2R^{(0)}\right)\right\}
 +\Half
 \Tr_V\ln\left\{Z_{Nk}\left(-\DAL-\frac{R}{4}+k^2R^{(0)}\right)\right\}
\nnn \cm
 -\Half
 \Tr_0\ln\left\{Z_{Nk}\left(-\DAL-\frac{R}{2}+k^2R^{(0)}\right)\right\}
+\Half
\Tr_V\ln\left\{Z_{Nk}\left(-\DAL+\frac{R}{4}+k^2R^{(0)}\right)\right\}
\nnn \cm
+\Half  \Tr_0\ln\left\{Z_{Nk}\left(-\DAL+\frac{A_1+\sqrt{A_1^2-4B_1}}{2}
+k^2R^{(0)}\right)\right\}
\nnn \cm
+\Half  \Tr_0\ln\left\{Z_{Nk}\left(-\DAL+\frac{A_1-\sqrt{A_1^2-4B_1}}{2}
+k^2R^{(0)}\right)\right\}
\nnn \cm
-\Half  \Tr_V\ln\left\{\left(-\DAL+\frac{R}{4}+k^2R^{(0)}\right)\right\}
-\Tr_V\ln\left\{\left(-\DAL-\frac{R}{4}+k^2R^{(0)}\right)\right\}
\nnn \cm
-\Tr_0\ln\left\{\left(-\DAL+\frac{R}{K-3}+k^2R^{(0)}\right)\right\}
+\Tr_0\ln\left\{\left(-\DAL-\frac{R}{2}+k^2R^{(0)}\right)\right\}
							\label{(3.18)}
\ear
where
\[
	A_1=\frac{R}{K-3}+m_0^2  , \inch  B_1=\frac{4m_0^2\Lambda}{K-3}
\]
and $\Lambda$ should be replaced  by $\bar{\lambda}_k$ in (3.18).
Thus we have got the average effective action.

\section{Evolution equations for the Newtonian and cosmological constants}

 In this section we write down the renormalization group equation (2.3) for
 the action (3.1). The l.h.s. of the truncated evolution equation is given by
 (2.6), where we have projected the evolution equation on the space with low
 derivatives.

 Now we want to find the r.h.s. of the evolution equation. To
 this end, we differentiate \eq (3.18) with respect to $t$. Then we
 expand the operators in (3.18) in the curvature $R$ because we
 are only interested in terms of the order $\int d^4x \sqrt{g}$ and $\int
 d^4x \sqrt{g}R$. We also have to expand some functions of $R$ inside the
 operators that appear in the first four terms and in the eighth and ninth
 terms in \eq (3.18) up to linear terms in $R$.

 Let us take the first two terms in \eq (3.18) and represent
\beq
  	f_{1,2}(R)=\frac{b\pm \sqrt{b^2-4c}}{2}   	\label{(4.1)}
\eeq
where $b$ and $c$ are given by (3.17). These functions may be written as
\beq
    f_{1,2}(R)=\Half \left[\frac{R}{6}+A\pm\sqrt{BR^2+CR+D}\right]
                                                        \label{(4.2)}
\eeq
 where $A$,$B$,$C$,$D$ depend on $f$,$\nu$,$\Lambda$ through $b$, $c$,
 $m_0^2$ and $m_2^2$. Expanding (4.2) up to terms linear in curvature
\beq
f_{1,2}=\Half \left[\frac{R}{6}+A\pm\left(D^{1/2}+\Half
D^{-1/2}\left(BR^2+CR\right)+ ...\right)\right]
\simeq\Half \left[A\pm D^{1/2}+R\left(\frac{1}{6}\pm\Half
CD^{-1/2}\right)\right],
							\label{(4.3)}
\eeq
we can write
\[
	f_1=\alpha_1 R+\alpha_2, \inch  f_2=\beta_1 R+\beta_2
\]
where
\bearr
\alpha_1,\ \beta_1=\frac{1}{4}\left\{\frac{1}{3}\pm CD^{-1/2}\right)=
	\frac{1}{12}+\frac{f^2+\nu^2}{6\nu^2}\pm\Half
	\left(\frac{f^2+\nu^2}{3\nu^2}-\frac{K+6}{6K}\right)m_2^2
	\left[m_2^4+\frac{8\Lambda m_2^2}{K}\right]^{-1/2};
\nnn
	\alpha_2,\ \beta_2=\Half \left(A\pm D^{1/2}\right)
 	=\Half m_2^2\pm\Half \left[m_2^4+\frac{8\Lambda
 				  m_2^2}{K}\right]^{1/2}    \label{(4.4)}
\ear
 Here we have replaced the values of $C$ and $D$.

The functions $f_1$ and $f_2$ are the same for the third
and fourth terms in (3.18). In a similar manner we obtain for
the eighth and ninth terms in (3.18) the following values:
\[
	f_3=\gamma_1 R+\gamma_2, \inch   f_4=\delta_1R+\delta_2
\]
with
\bearr
	\gamma_1,\ \delta_1=\frac{1}{2(K-3)}\pm
    \frac{m_0^2}{2(K-3)}\left(m_0^4-\frac{16m_0^2\Lambda}{K}\right)^{-1/2},
\cm
	\gamma_2,\ \delta_2=\Half m_0^2\pm \Half \left(m_0^4
		-\frac{16m_0^2\Lambda}{K-3}\right)^{1/2}.
							   \label{(4.5)}
\ear
 After this expansion we can expand the operators in (3.18) linearly in $R$.
 Let us introduce the notation
\beq
	\Delta_i^{-1}\left(\alpha R+\lambda+k^2R^{(0)}\right)
	=\Delta_i^{-1}\left(\lambda+k^2R^{(0)}\right)-
	\Delta_i^{-2}\left(\lambda+k^2R^{(0)}\right)\alpha R+0(R^2)
							\label{(4.6)}
\eeq
 where $a$ takes the values $\alpha_1,\ \beta_1,\ \gamma_1,\ \delta_1$ and
 $\lambda$ takes the values $\alpha_2,\ \beta_2,\ \gamma_2,\ \delta_2$,
 given by (4.4), (4.5). For a more compact notation let us introduce
\bearr
	\Delta_{i\lambda}(z)=\Delta_i\left(\lambda+k^2R^{0}(z)\right),
				       		\label{(4.7)}
\\ \lal
	N(z)=\frac{\d_t [Z_{Nk}k^2R^{(0)}(z) }{Z_{Nk}}, \cm
	N_0(z)=\d_t\left[k^2R^{(0)}(z)\right].
						\label{(4.8)}
\ear
 Here the variable $z$ replaces $-\DAL/k^2$. Note that we use as a cut-off
 the same function as in \Ref{mr}:
 $R^{(0)}(z)=z/ {\rm Exp}\,[z]-1)$. The above steps then lead to
\bearr
\d_t\Gamma_k[g,g]=\Half \Tr_T\left[N\Delta_{T\alpha_2}^{-1}\right]
	         +\Half \Tr_T\left[N\Delta_{T\beta_2}^{-1}\right]-
	\Half \Tr_V\left[N\Delta_{V\alpha_2}^{-1}\right]
	-\Half \Tr_V\left[N\Delta_{V\beta_2}^{-1}\right]
	      +\Tr_V\left[N\Delta_{V0}^{-1}\right]
\nnn \inch
	-\Half \Tr_S\left[N\Delta_{S0}^{-1}\right]
	+\Half \Tr_S\left[N\Delta_{S\gamma_2}^{-1}\right]
	+\Half \Tr_S\left[N\Delta_{S\delta_2}^{-1}\right]
	-\frac{3}{2}\Tr_V\left[N_0\Delta_{V0}^{-1}\right]
\nnn \inch
	-R\biggl\{\Half \alpha_1\Tr_T\left[N\Delta_{T\alpha_2}^{-2}\right]
	+\Half \beta_1\Tr_T\left[N\Delta_{T\beta_2}^{-2}\right]
  	-\Half \left(\alpha_1-\frac{5}{12}\right)\Tr_V
	\left[N\Delta_{V\alpha_2}^{-2}\right]
\nnn \inch
	-\Half \left(
	\beta_1-\frac{5}{12}\right)\Tr_V\left[N\Delta_{V\beta_2}^{-2}\right]
  +\frac{1}{4}\Tr_S\left[N\Delta_{S0}^{-2}\right]+\Half \gamma_1
	\Tr_S\left[N\Delta_{S\gamma_2}^{-2}\right]
\nnn \inch
	+\Half \delta_1\Tr_S \left[N\Delta_{S\delta_2}^{-2}\right]
  	+\frac{1}{8}\Tr_V\left[N_0\Delta_{V0}^{-2}\right]
	-\left(\frac{1}{K-3}+\Half \right)\Tr_S\left[N_0
	\Delta_{S0}^{-2}\right]\biggr\}.                    \label{(4.9)}
\ear
The terms with $N_0$ are contributions of the ghosts.

As a next step we evaluate the traces. We use the heat kernel expansion
which for an arbitrary function of the covariant Laplacian $W(D^2)$ reads
\beq
\Tr_j [W(-D^2)] =
    (4\pi)^{-2}\tr_j(I)\left\{Q_2[W]\int d^4x\sqrt{g}
    +\frac{1}{6}Q_1[W]\int d^4x\sqrt{g}R+0(R^2)\right\}    \label{(4.10)}
\eeq
 where by $I$ we denote the unit matrix in the space of fields on which
 $D^2$ acts. Therefore $\tr_j(I)$ simply counts the number of independent
 degrees of freedom of the field, namely
\[
     \tr_s(I)=1, \cm   tr_V(I)=4, \cm  tr_T(I)=9.
\]
 The sort $j$ of fields enters into (4.6) via $\tr_j(I)$ only. Therefore we
 will drop the index $j$ of $\Delta_{ja}$ after evaluation of the traces
 in the heat kernel expansion.

 The functionals $Q_n$ are Mellin transforms of $W$:
\beq
	Q_0[W]=W[0], \inch
    Q_n[W]=\frac{1}{\Gamma(n)}\int_0^{\infty}dz z^{n-1}W(z) \cm
                   (n>0).                                    \label{(4.11)}
\eeq

 Now we have to perform the heat kernel expansion (4.10) in \eq (4.11). This
 leads to a polynomial in $R$ which is the r.h.s. of the evolution equation.
 By comparison of coefficients with the l.h.s. of the evolution
 equation (2.6) we obtain in the order of $\int d^4x\sqrt{g}$
\bearr
\d_t [Z_{Nk}\Lambda_k] =
	\frac{1}{8\kappa^2}\frac{1}{(4\pi)^2}
	\biggl\{5Q_2\left[N\Delta_{\alpha_2}^{-1}\right]+5Q_2
	\left[N\Delta_{\beta_2}^{-1}\right]+7Q_2
		\left[N\Delta_0^{-1}\right]
\nnn \inch
  +Q_2\left[N\Delta_{\gamma_2}^{-1}\right]+Q_2
	\left[N\Delta_{\delta_2}^{-1}\right]-6Q_2\left[N_0
		\Delta_0^{-1}\right]\biggr\}                \label{(4.12)}
\ear
	And in the order of $\int d^4x \sqrt{g}R$
\bearr
		\d_t(Z_{Nk})=-\frac{1}{24\kappa^2}\frac{1}{(4\pi)^2}
	\biggl\{5Q_1\left[N\Delta_{\alpha_2}^{-1}\right]+5Q_1\left[N
		\Delta_{\beta_2}^{-1}\right]
  	+7Q_1\left[N\Delta_0^{-1}\right]
  	          +Q_1\left[N\Delta_{\gamma_2}^{-1}\right]
\nnn \inch
	+Q_1\left[N\Delta_{\delta_2}^{-1}\right]-12Q_1\left[N_0
		\Delta_0^{-1}\right]
  	-30\left(\alpha_1+\frac{1}{12}\right)Q_2\left[N
	\Delta_{\alpha_2}^{-2}\right]-30\left(\beta_1+\frac{1}{12}\right)
	Q_2\left[N\Delta_{\beta_2}^{-2}\right]
\nnn \inch
  	-3Q_2\left[N\Delta_0^{-2}\right]-6\gamma_1 Q_2\left[N
	\Delta_{\gamma_2}^{-2}\right]
  	-6\delta_1Q_2\left[N\Delta_{\delta_2}^{-2}\right]
  +\frac{12}{K-3}Q_2\left[N_0\Delta_0^{-2}\right]\biggr\}.     \label{(4.13)}
\ear

	Let us introduce the cut-off-dependent integrals
\bearr
	\Phi_n^p(w)=\frac{1}{\Gamma(n)}\int_0^{\infty}dz z^{n-1}
	\frac{R^{(0)}(z)-zR^{(0)'}(z)}{[z+R^{(0)}(z)+w]^p},
\nnn
	\tilde{\Phi}_n^p(w)=\frac{1}{\Gamma(n)}\int_0^{\infty}dz
	z^{n-1}\frac{R^{(0)}(z)}{[z+R^{(0)}(z)+w]^p}        \label{(4.14)}
\ear
for $n>0$. It follows that $\Phi_0^p(w)=\tilde{\Phi}_0^p(w)=(1+w)^{-p}$ for
$n=0$. In addition, we use the fact that
\beq
	N=\frac{\d_t\left[Z_{Nk}k^2R^{(0)}(-D^2/k^2)\right]}
	{Z_{Nk}}=\left[2-\eta_N(k)\right]k^2R^{(0)}(z)+2D^2R^{(0)'}(z)
							\label{(4.15)}
\eeq
 with $\eta_N(k)=-\d_t(\ln Z_{Nk})$ being the anomalous dimension
 of the operator $\sqrt{g}R$. Then we can rewrite \eqs (4.8) and
 (4.9) in terms of $\Phi$ and $\tilde{\Phi}$. This leads to the following
 set of equations:
\bearr
 \d_t [Z_{Nk}\Lambda_k]
      		=\frac{1}{4\kappa^2}\frac{1}{(4\pi)^2}k^4
 \biggl\{5\Phi_2^1(\alpha_{2k})+5\Phi_2^1(\beta_{2k})-5\Phi_2^1(0)
  +\Phi_2^1(\gamma_{2k})+\Phi_2^1(\delta_{2k})
\nnn \cm\cm
  	-\Half \eta_N(k)
  \left[5\tilde{\Phi}_2^1(\alpha_{2k})+5\tilde{\Phi}_2^1(\beta_{2k})
   +7\tilde{\Phi}_2^1(0)+\tilde{\Phi}_2^1(\gamma_{2k})
	+\tilde{\Phi}_2^1(\delta_{2k})\right]\biggr\},      \label{(4.16)}
\\ \lal
	\d_t(Z_{Nk})=
			-\frac{1}{24\kappa^2}\frac{1}{(4\pi)^2}k^2
      \biggl\{10\Phi_1^1(\alpha_{2k})+10\Phi_1^1(\beta_{2k})
 	-10\Phi_1^1(0)+2\Phi_1^1(\gamma_{2k})+2\Phi_1^1(\delta_{2k})
\nnn \cm \cm
	-(60\alpha_1+5)\Phi_2^2(\alpha_{2k})-(60\beta_1+5)
		\Phi_2^2(\beta_{2k})
  	+\biggl(\frac{24}{K-3}-6\biggr)\Phi_2^2(0)
  	-12\gamma_1\Phi_2^2(\gamma_{2k})-12\delta_1\Phi_2^2(\delta_{2k})
\nnn \cm    \cm
  -\eta_N(k)\biggl[5\tilde{\Phi}_1^1(\alpha_{2k})+5\tilde{\Phi}_1^1
	(\beta_{2k})+7\tilde{\Phi}_1^1(0)
	+\tilde{\Phi}_1^1(\gamma_{2k})
   	+\tilde{\Phi}_1^1(\delta_{2k})-30(\alpha_1+\fract{1}{12})
		\tilde{\Phi}_2^2(\alpha_{2k})
\nnn \cm       \cm
	-30(\beta_1+\fract{1}{12})\tilde{\Phi}_2^2 (\beta_{2k})
   -3\tilde{\Phi}_2^2(0)-6\gamma_1\tilde{\Phi}_2^2(\gamma_{2k})
	-6\delta_1\tilde{\Phi}_2^2(\delta_{2k})\biggr]\biggr\}\label{(4.17)}
\ear
	with
\bear
\alpha_{2k} \eql \alpha_2/k^2=f^2\kappa_k^2+f^2\kappa_k^2
\left(1+\frac{4\lambda_k}{Kf^2\kappa_k^2}\right)^{1/2}   ,
\nn
\beta_{2k} \eql \beta_2/k^2=f^2\kappa_k^2-f^2\kappa_k^2\left(1
+\frac{4\lambda_k}{Kf^2\kappa_k^2}\right)^{1/2}   ,
\nn
\gamma_{2k} \eql \gamma_2/k^2=\nu^2\kappa_k^2+\nu^2\kappa_k^2
\left(1-\frac{8\lambda_k}{(K-3)\nu^2\kappa_k^2}\right)^{1/2}    ,
\nn
\delta_{2k} \eql \delta_2/k^2=\nu^2\kappa_k^2-\nu^2\kappa_k^2
\left(1-\frac{8\lambda_k}{(K-3)\nu^2\kappa_k^2}\right)^{1/2},  \label{(4.18)}
\ear
where $\kappa_k^2=\kappa^2/k^2$ and $\lambda_k=\Lambda/k^2$ and
we have replaced the values of $m_2^2$ and $m_0^2$ in \eqs (4.4)
and (4.5), respectively.
Now we introduce the dimensionless, renormalized Newtonian constant
\beq
		g_k=k^2G_k=k^2Z_{Nk}^{-1}\bar{G}.   \label{(4.19)}
\eeq
 Here $G_k$ is the renormalized Newtonian constant at scale $k$. The
 evolution equation for $g_k$ then reads
\beq
\d_t g_k=\left[2+\eta_N(k)\right]g_k.                \label{(4.20)}
\eeq
>From (4.18) we find the anomalous dimension $\eta_N(k)$:
\beq
\eta_N(k)=g_kB_1(\kappa_k,\lambda_k)+\eta_N(k)g_kB_2(\kappa_k,\lambda_k)
							\label{(4.21)}
\eeq
	where
\bearr
	B_1(\kappa_k,\lambda_k)
		=\frac{1}{12\pi}\biggl\{10\Phi_1^1(\alpha_{2k})
		+10\Phi_1^1(\beta_{2k})
 	-10\Phi_1^1(0)+2\Phi_1^1(\gamma_{2k})+2\Phi_1^1(\delta_{2k})
	-(60\alpha_1+5)\Phi_2^2(\alpha_{2k})
\nnn \inch
	-(60\beta_1+5) \Phi_2^2(\beta_{2k})
   +\biggl(\frac{24}{K-3}-6\biggr)\Phi_2^2(0)-12\gamma_1\Phi_2^2(\gamma_{2k})
	-12\delta_1\Phi_2^2(\delta_{2k})\biggr\},
\nnn
		B_2(\kappa_k,\lambda_k)=-\frac{1}{12\pi}
	\biggl\{5\tilde{\Phi}_1^1(\alpha_{2k})+5\tilde{\Phi}_1^1
	(\beta_{2k})+7\tilde{\Phi}_1^1(0)
 	+\tilde{\Phi}_1^1(\gamma_{2k})+\tilde{\Phi}_1^1
	(\delta_{2k})-30(\alpha_1+\fract{1}{12})\tilde{\Phi}_2^2(\alpha_{2k})
\nnn \inch
	-30(\beta_1+\fract{1}{12})\tilde{\Phi}_2^2(\beta_{2k})
  -3\tilde{\Phi}_2^2(0)-6\gamma_1\tilde{\Phi}_2^2(\gamma_{2k})
	-6\delta_1\tilde{\Phi}_2^2(\delta_{2k})\biggr\}.       \label{(4.22)}
\ear
	Solving (4.21),
\beq
	\eta_N(k)=\frac{g_kB_1(\kappa_k,\lambda_k)}
		{1-g_kB_2(\kappa_k,\lambda_k)},                \label{(4.23)}
\eeq
 we see that the anomalous dimension $\eta_N$ is a nonpertubative
 quantity. From (4.16) we obtain the evolution equation for the cosmological
 constant
\bearr
	\d_t(\lambda_k)=-\left[2-\eta_N(k)\right]\lambda_k+\frac{g_k}{4\pi}
	\biggl\{10\Phi_2^1(\alpha_{2k})+10\Phi_2^1(\beta_{2k})
	-10\Phi_2^1(0)
	+2\Phi_2^1(\gamma_{2k})+2\Phi_2^1(\delta_{2k})
\nnn \inch
	-\eta_N(k)
	\left[5\tilde{\Phi}_2^1(\alpha_{2k})+5\tilde{\Phi}_2^1(\beta_{2k})
  	+7\tilde{\Phi}_2^1(0)+\tilde{\Phi}_2^1(\gamma_{2k})
	+\tilde{\Phi}_2^1(\delta_{2k})\right]\biggr\}.    \label{(4.24)}
\ear
 \eqs(4.20) and (4.24) with (4.23) determine the value of the running
 Newtonian constant and cosmological constant at the scale
 $k<<\Lambda_{\rm cut-off}$. The above evolution equations include
 nonperturbative effects which go beyond a simple one-loop calculation. This
is particularly obvious if one expands (4.23) for small values of $g_k$:
\beq
 \eta_N=g_kB_1(\kappa_k,\lambda_k)\Bigl[1+g_kB_2(\kappa_k,\lambda_k)
 +g_k^2B_2^2(\kappa_k,\lambda_k)+...\Bigr].                 \label{(4.25)}
\eeq

\section{Critical points and the running Newtonian coupling constant}

In the present section we give some remarks about the properties of
nonperturbative RG equations. First of all, let us estimate the qualitative
behaviour of the running gravitational coupling constant.

The dimensional quantity $G_k$ evolves according to
\beq
	\d_tG_k=\eta_NG_k                  \label{(5.1)}
\eeq
The set of RG equations for the coupling constants is too complicated and
cannot be solved analytically. Hence we assume that the cosmological
constant is much smaller than the IR cut-off scale, $\lambda_k<<k^2$, so we
can put $\lambda_k\sim 0$. This simplifies \eqs (4.29), (4.24) and we are
left with only \eq (5.1). After that we perform an expansion in powers of
$(\bar{G}_k^2)^{-1}$ keeping only the first term (i.e. we evaluate the
functions $\Phi_n^p(0)$ and $\bar{\Phi}_n^p(0)$) and finally obtain (with
$g_k\sim k^2\bar{G}$)
\beq
	G_k=G_o\left[1-w\bar{G}k^2+...\right]      \label{(5.2)}
\eeq
	where
\[
	w=-\Half B_1(0,0)=\frac{1}{24\pi}\left[\left(50+22\frac{f^2}{\nu^2}
		\right)-\frac{7\pi^2}{3}\right].
\]
In obtaining $w$ we use the same cut-off function as in \Ref{mr}.

For Einstein gravity in the same formalism (also using the gauge-fixing
independent EA) we have got [9]:
\beq
	w=\frac{\pi}{36}\left[\frac{108}{\pi^2}-1\right]  \label{(5.3)}
\eeq
 In the case under discussion we see that the sign of $w$ depends on
 the higher-derivative coupling constants:
\beq
	w>0  \cm \mbox{if}  \cm 50-\frac{7\pi^2}{3}+\frac{22f^2}{\nu^2}>0.
		\label{(5.4)}
\eeq
\twocol
\noi
 The coupling constant $\nu^2$ maybe chosen to be negative (see
 \cite{bos}). So, e.g., for $f^2=1$, $\nu^2=\pm 1$ we get $w>0$ and
 the Newtonian coupling decreases as $k^2$ increases. In other words, we find
 an antiscreening behaviour of the gravitational coupling. On the
 contrary, for $f^2=1$, $\nu^2=-1/2$ we get $w<0$ and a screening behaviour
 for the Newtonian coupling (for the one-loop behaviour of the Newtonian
 coupling in $R^2$-gravity with matter, see \cite{el}). This means that in
 such a phase the gravitational charge (mass) is screened by quantum
 fluctuations, or, in other words, the Newtonian coupling is smaller at
 smaller distances. The sign of a quantum correction to the Newtonian
 potential will be different too. Note that our solution (5.2) is actually
 qualitative, and the full RG system should be analyzed for a better result.

 Our main qualitative result is that $R^2$-gravity considered as an
 effective theory may change the low-energy gravitational phenomena as
 compared with Einstein gravity.

 Let us now investigate the problem of
 existence of critical points in the theory under study. We search for
 points at which the  r.h.s. of \eqs (4.20) and (4.24) are equal to zero
 (supposing $g_k=k^2\bar{G}$). For Einstein gravity such a study may be
 carried out quite easily (since the functions $B_1$ and $B_2$ depend only
 on $\lambda_k$). A numerical analysis of the corresponding RG system which
 is written in the physical Landau-DeWitt gauge gives \Ref{fo}:
 \beq
	\lambda_k=0.352, \cm     g_k=0.348.            \label{(5.5)}
\eeq
 These points actually correspond to UV-stable fixed points. Note that
 the solutions (5.5) do not give a solution to the cosmological constant
 problem as a result of the non-perturbative RG behaviour.

 In $R^2$-gravity the
situation is much more complicated because the functions $B_1$ and $B_2$
depend on $\kappa_k$ and $\lambda_k$ and because there are the
higher-derivative coupling constants as free parameters of the theory.
Supposing $g_k=k^2\bar{G}$, we can get the following unstable fixed points
for $f^2=1/10$ and $\nu^2=-1/4$:
\beq
	\lambda_k=4.47 , \cm    g_k=4.46. \label{(5.6)}
\eeq

Note that for other values of the higher-derivative coupling constants
one can get numerically other values for the unstable fixed points.

\section{Discussion}

In the present work we have studied the truncated evolution
equation in higher-derivative quantum gravity. Making a truncation to the
space of low-derivative functionals, we have obtained nonperturbative RG
equations for the Newtonian and cosmological coupling constants. The
necessary step in such a study is the calculation of the effective average
action on some background (we have used the de Sitter background).

The properties of the nonperturbative RG equations (like the existence of
critical points and the behaviour of the Newtonian coupling) are discussed.
As we showed, the higher-derivative QG may behave at low energies
qualitatively different from Einstein QG.

The next open problem in such an approach is to make a better truncation of
the evolution equation to the space of functionals with higher derivatives.
In such a way one could obtain a complete set of nonperturbative RG
equations for all coupling constants:  $f^2$, $\nu^2$, $\kappa^2$,
$\Lambda$. Hence, unlike the present study where $f^2$ and $\nu^2$ are free
parameters, we might define the critical points of the complete phase space
(the RG equations for $\kappa^2$ and $\Lambda$ are, of course, the same).

 However, in order to find the effective average action with a truncation to
 the space of higher-derivative functionals, we have to perform a
 calculation of the one-loop effective action in a background where $R^2$
 and $C_{\mu\nu\alpha\beta}^2$ may be distinguished. Clearly the de Sitter
 space does not belong to this class of backgrounds.

 As far as we know (see \cite{bos} for a review), the one-loop effective
 action for $R^2$-gravity has been found only in the de Sitter or flat
 backgrounds. Even such a calculation is extremely complicated. A
 generalization of such a result to a more complicated background (say, of
 the above sort) being, in principle, possible, is extremely complicated.
 Moreover, that is just one step in writing the r.h.s. of the evolution
 equation. After that, much more work is required to obtain explicitly the
 nonperturbative  RG equations for $\nu^2$ and $f^2$. Hence, this problem is
 left for future research.

Another related problem is the gauge dependence of the average effective
action. In order to solve this problem, one has to do an even better
truncation which includes all gauge parameters as independent functions of
$k$. Hence, in addition to the four RG equations for the coupling constants,
one should write some RG equations for all gauge parameters. Then the RG
equations for the gauge parameters should lead to some stable fixed points.
These fixed point values for the gauge parameters should be used in the RG
equations for the coupling constants.  It is clear that such a programme is
too complicated and cannot be realized.

 However, there is a simpler way which we have actually used in this work.
 In a study of the effective average action for the Yang-Mills theory (see
 \cite{ull}) a $k$-dependent gauge parameter was used. It has been shown
 that there exists an attracting fixed point of the truncated evolution for
 the gauge parameter. This fixed point corresponds to the so-called
 Landau-DeWitt gauge.  Hence, the alternative easy way of studying the
 truncated evolution equation is to work in the physical Landau-DeWitt
 gauge (actually it corresponds to a study in the gauge-fixing-independent
 effective action formalism). Similarly in Einstein gravity, in order to
 avoid the introduction of a tedious additional RG equation for the gauge
 parameter, one can work in the Landau-DeWitt gauge which again corresponds
 to the gauge-fixing-independent EA formalism (see \cite{fo}).  In the same
 way, we have used here the gauge-fixing-independent EA in order to solve
 the gauge dependence problem for $R^2$-gravity in our formulation.

 Hence, our study which indicates the
 qualitative  difference between Einstein and $R^2$-gravity even at low
 energy scales is a necessary step in the formulation of better
 truncations of the evolution equation in $R^2$-gravity. Moreover, it is
 expected to be useful also in the studies of supersymmetric
 $R^2$-gravity in a nonperturbative approach.

\Acknow
{We would like to thank A.A. Bytsenko for helpful discussions and
 participation in part of this work and A. Romeo for help in  numerical
 calculations. We are grateful to M. Reuter for useful e-mail discussions.
 L.N.G. was supported by COLCIENCIAS (Colombia) Project No. 1106-05-393-95.
 S.D.O was supported in part by COLCIENCIAS.
 }

\end{document}